# A core-shell-surface layer model to explain the size dependence of effective magnetic anisotropy in magnetic nanoparticles

Sobhit Singh, Kelly L. Pisane, and Mohindar S. Seehra*

*Department of Physics and Astronomy, West Virginia University, Morgantown, WV-26506, USA*

*Corresponding author:* Mohindar.Seehra@mail.wvu.edu

*Abstract*—The particle size (*D*) dependence of the effective magnetic anisotropy $K_{eff}$ of magnetic nanoparticles (NPs) usually shows $K_{eff}$ increasing with decreasing *D*. This dependence is often interpreted using the Eq.: $K_{eff} = K_b + (6K_S/D)$ where $K_b$ and $K_S$ are the anisotropy constants of the spins in the bulk-like core and surface layer, respectively. Here, we show that this model is inadequate to explain the observed size-dependency of $K_{eff}$ for smaller nanoparticles with *D* < 5 nm. Instead the results in NPs of maghemite (γ-Fe$_2$O$_3$), NiO and Ni are best described by an extension of the above model leading to the variation given by $K_{eff} = K_b + (6K_S/D) + K_{sh}\{[1-(2d/D)]^{-3} -1\}$, where the last term is due to the spins in a shell of thickness *d* with anisotropy $K_{sh}$. The validation of this core-shell-surface layer (CSSL) model for three different magnetic NPs systems viz. ferrimagnetic γ-Fe$_2$O$_3$, ferromagnetic Ni and antiferromagnetic NiO suggests its possible applicability for all magnetic nanoparticles.

## I. INTRODUCTION

The current on-going interest in the properties of magnetic nanoparticles (NPs) is primarily due to two reasons: (i) The magnetic properties of NPs are strongly dependent on their size *D* because of the increasing role of the surface spins whose concentration increases with decreasing *D* as 1/*D*; and (ii) Magnetic NPs have diverse applications in many areas such as magnetic storage media, biomedicine, magnetic drug delivery, sensors, ferrofluids and catalysts [1-5]. An important property of magnetic NPs is the blocking temperature $T_B$ which is related to the effective magnetic anisotropy $K_{eff}$ and volume *V* of the NPs through the relation [6-8]:

$$T_B = T_0 + \frac{K_{eff}V}{k_B \ln\left(\frac{f_0}{f_m}\right)} \quad (1)$$

Here $k_B$ is the Boltzmann constant, $f_0 \sim 10^{10} - 10^{12}$ Hz is the system-dependent attempt frequency varying only weakly with temperature, $f_m$ is the experimental measurement frequency and $T_0$ is an effective temperature representing the strength of the interparticle interactions (IPI). The importance of the anisotropy energy $K_{eff}V$ is that it keeps the magnetic moment of the NP aligned in a particular direction. Therefore, how $K_{eff}$ varies with the size *D* or volume *V* of the NPs is important with regard to the stored information in the recorded media. To determine $K_{eff}$ using Eq. (1) for a particle of volume *V*, $T_B$ needs to be measured at several frequencies $f_m$ which then allows determination of $T_0$, $K_{eff}$ and $f_0$ for the system. [9-10] Determining $K_{eff}$ from the measured $T_B$ assuming $T_0 = 0$ (*i.e.* no IPI) and undetermined $f_o$ value leads to error in the magnitude of $K_{eff}$. To reduce the strength of the IPI which can include dipole-dipole and exchange interactions, the NPs are often coated with surfactants or diluted in diamagnetic hosts. [8-15]. In a recent paper [10], we reported on the $K_{eff}$ vs. *D* variation in the maghemite (γ-Fe$_2$O$_3$) NPs in the size range *D* = 2.5 nm to 15 nm including some data taken from the literature and showed that this variation is not adequately described by the current core-surface layer model [16]. Instead a core-shell-surface layer (CSSL) model was proposed to explain the observed unusual enhancement of $K_{eff}$ for the smaller sizes [10]. In this paper, we further elaborate on this CSSL model and also test its general validity for the NPs of antiferromagnetic NiO and ferromagnetic Ni. Details of the relevant important issues and results are presented in the following sections.

## II. INTERPARTICLE INTERACTIONS (IPI)

It is evident from Eq. (1) that $T_o$ describing the strength of IPI in a given system needs to be determined first for accurate determination of $K_{eff}$ for the system. For this, $T_B$ needs to be measured at least for two sufficiently different frequencies $f_m$ followed by calculating the quantity $\Phi$ defined as follows [7, 8, 11]:

$$\Phi = \frac{T_B(2) - T_B(1)}{T_B(1)[\log f_m(2) - \log f_m(1)]} \quad (2)$$

Here $T_B(1)$ and $T_B(2)$ are the blocking temperature measured at two sufficiently different frequencies $f_m(1)$ and $f_m(2)$. For no IPI ($T_0 = 0$), $\Phi \sim 0.13$ and for $\Phi < 0.13$, the magnitude of IPI and $T_0$ increases with decreasing magnitude of $\Phi$. Experimentally, the ac magnetic susceptibility is an ideal way to probe the frequency dependence of $T_B$, where $T_B$ is best determined by the peak position of the χ" *vs.* temperature data, with χ" being the out-of-phase component of the ac-susceptibility. For magnetic NPs, $T_B$ shifts to higher temperatures with increase in $f_m$ as shown in Fig. 1 for the oleic-acid coated 6.3 nm diameter NPs of maghemite [9]. Analysis of this data shows that $\Phi = 0.084$, $T_0 = 11$ K, $f_0 =$

2.6 × $10^{10}$ Hz, and $K_{eff}$ = 7.5 × $10^5$ erg/cm$^3$. [9, 10]. In Fig. 2, the fits of the data for the maghemite NPs of size $D$ = 2.5, 3.4, 6.3, and 7.0 nm to Eq. (1) from our investigations [9] are shown to emphasize the importance of determining $T_0$ before evaluating $K_{eff}$.

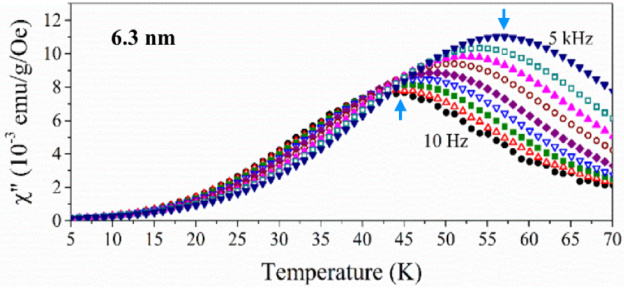

Figure 1: Temperature dependence of the imaginary part of ac magnetic susceptibilities ($\chi''$) data for 6.3 nm diameter $\gamma$-Fe$_2$O$_3$ NPs, measured at frequencies $f_m$ =10 Hz, 20 Hz, 50 Hz, 100 Hz, 200 Hz, 500 Hz, 1 kHz, 2 kHz, and 5kHz. The arrows mark the peak positions of $\chi''$ defining the blocking temperature $T_B$ at 10 Hz and 5 kHz frequencies.

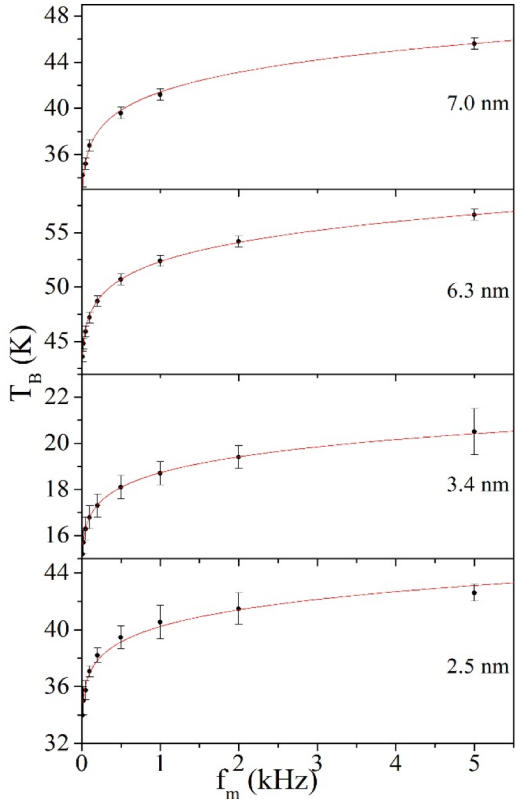

Figure 2: Variation of the blocking temperature $T_B$ with measuring frequency $f_m$ for four $\gamma$-Fe$_2$O$_3$ NPs of size $D$ = 2.3, 3.4, 6.3, and 7.0 nm [9]. The solid lines are fits to Eq. (1) using $f_o$ = 2.6 × $10^{10}$ Hz and $T_o$ = 0, 11, 2.5 and 12.5 K K for the $D$ = 7.0, 6.3, 3.4, and 2.5 nm size NPs, respectively.

## III. THE CORE-SURFACE LAYER MODEL

The spins in the core of a magnetic NP usually have bulk-like magnetic ordering with bulk-like anisotropy $K_b$. However, the spins on the surface of NPs experience different anisotropy $K_S$ because of the broken crystalline symmetry and broken exchange bonds at the surface. Considering the surface effects, Bødker et al. [16] proposed the following equation to explain the size dependence of $K_{eff}$ in Fe NPs:

$$K_{eff} = K_b + \frac{6 K_S}{D} \qquad (3)$$

Here the factor $6/D$ is the ratio of the surface area to volume of a spherical NP with diameter $D$. This relationship is often quoted in the literature but in some cases deviations from this variation have also been reported. [10, 12, 17, 18]

In a recent paper [10], we have tested the validity of Eq. (3) for a total of 18 maghemite ($\gamma$-Fe$_2$O$_3$) NPs in the size range of 2.5 nm to 15 nm. The data included four NPs with $D$ = 2.3, 3.4, 6.3, and 7.0 nm from our own investigations [9] along with the available data of the remaining 14 NPs taken from the literature [10]. Care was taken to select only those data points for which the effects of IPI were taken into account before determining $K_{eff}$ using Eq. (1). The plot of $K_{eff}$ vs. $1/D$ data for maghemite NPs (shown in Fig. 4) reveals that the expected linearity of the plot based on Eq. (3) is only valid for $D$ > 5 nm particles since for $D$ < 5 nm, there is an anomalous enhancement of $K_{eff}$ with decreasing $D$.

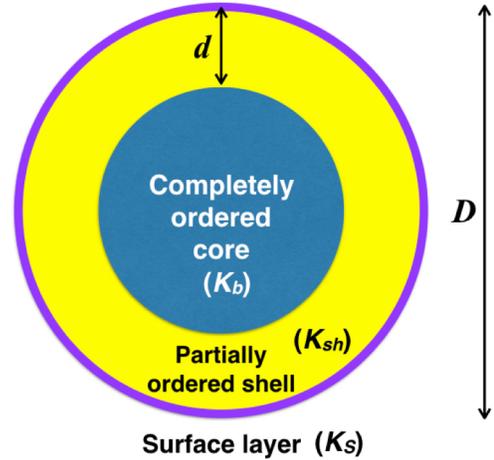

Figure 3: A pictorial representation of the core-shell-surface layer (CSSL) model for a NP of diameter $D$ with completely ordered spins in the core, partially ordered spins in a shell of thickness $d$ and the surface layer, with magnetic anisotropy values of $K_b$, $K_{sh}$ and $K_s$, respectively.

## IV. THE CORE-SHELL-SURFACE LAYER (CSSL) MODEL

To explain the observed deviations of the $K_{eff}$ vs. $1/D$ data for the smaller NPs of $\gamma$-Fe$_2$O$_3$ from the predictions of the core-surface layer model represented by Eq. (3), an extension of this model was proposed consisting of a core, a shell of thickness $d$ with anisotropy $K_{sh}$, and the usual atomically thin surface layer. A pictorial description of this core-shell-surface layer (CSSL) model is shown in Fig. 3 and it is described by the following Eq.: [10]

$$K_{eff} = K_b + \frac{6 K_S}{D} + K_{sh}\left\{\left(1 - \frac{2d}{D}\right)^{-3} - 1\right\} \quad (4)$$

Here $K_{sh}$, the anisotropy of the spins in the shell, is different from $K_s$ and $K_b$. A justification for the CSSL model is the recent Monte-Carlo simulations [19] which showed that the surface disorder in maghemite NPs propagates towards the inside of the NPs with decreasing size, thus, forming a shell layer of finite thickness in which the arrangement of spins is different from those in the core and surface layer. Additionally, the recent neutron scattering measurements and the theoretical investigations further confirm the formation of a shell layer in the small magnetic NPs. [20-21] The factor $\{[1-(2d/D)]^{-3} -1\}$ in Eq. (4) represents the ratio of the shell volume to the core volume i.e. $[D^3 – (D-2d)^3]/(D-2d)^3$. Thus, it is a measure of the fraction of the spins in the shell experiencing an effective anisotropy $K_{sh}$ which is different from both $K_S$ and $K_b$. The validity of Eq. (4) is limited to $D > 2d$ since only in this limit the NPs have a core.

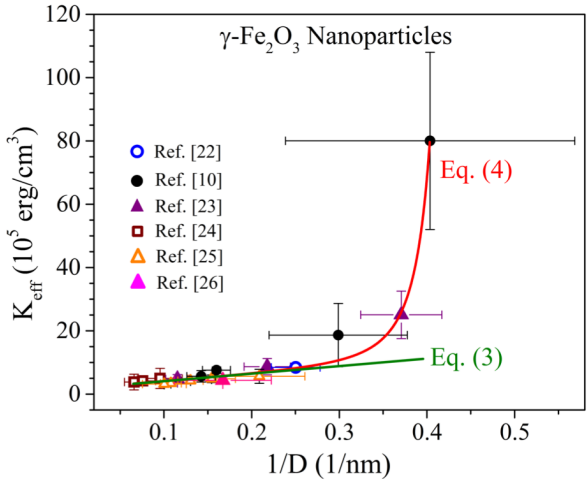

Figure 4: Variation of $K_{eff}$ with $1/D$ for maghemite NPs. Different symbols represent the data reported by different group as cited in the inset. The green line is the best fit corresponding to Eq. (3) while red line represents the fit corresponding to Eq. (4). The magnitudes of the fitting parameters are as follows: $K_b = 1.9 \times 10^5$ erg/cm$^3$, $K_S = 0.035$ erg/cm$^2$, $K_{sh} = 10570$ erg/cm$^3$, $d = 1.1$ nm.

The problem of determining the four fitting parameters $K_b$, $K_S$, $K_{sh}$, and $d$ of Eq. (4) can be simplified into two 2-parameters problems. We first determine the magnitudes of $K_b$ and $K_S$ from the linear fitting of Fig. 4 for $D > 5$ nm particles only, where we use Eq. (3) to fit the $K_{eff}$ vs. $1/D$ data. The other two parameters $K_{sh}$ and $d$ were then determined by taking the difference of the extrapolated linear curve to the smaller sizes and measured values of $K_{eff}$. The comparison of the data [10, 22-26] and the fitted curve for the whole range of 2.5 nm to 15 nm is shown in Fig. 4 with the magnitudes of the fitted parameters listed in the caption of the figure. The data of this figure is reproduced from our recent paper where details of the CSSL model were also presented. [10] Although, all the data points do not exactly fall on the fitted curve, as is often the case when comparing experimental data with theory, the overall trend of $K_{eff}$ vs. $D$ variation is well-captured by Eq. (4) within the experimental uncertainties.

## V. VALIDATION OF THE CSSL MODEL FOR OTHER SYSTEMS

In order to verify the general validity of the CSSL model, we next test its applicability for the NPs of antiferromagnetic NiO and ferromagnetic Ni using the data available in the literature. In selecting the data for these systems, again care was taken to include only those data points for which the effects of IPI have been taken into account before determining $K_{eff}$ [12, 13, 18, 27]. Measurements of $K_{eff}$ vs. $D$ for coated NiO NPs in which IPI was absent was reported by Shim et al. [12] where it was also shown that the variation of $K_{eff}$ vs. $D$ does not fit Eq. (3) due to the unusual enhancement of $K_{eff}$ for the smaller NPs [12]. This is similar to the observations reported here for $\gamma$-Fe$_2$O$_3$ NPs. The fit of the data in NiO NPs to the CSSL model of Eq. (4) is shown in the plot of $K_{eff}$ vs. $D$ in Fig. 5. Again the fitting of the $K_{eff}$ vs. $D$ data using Eq. (3) only captures variation for the larger NPs from which we estimated the $K_b$ and $K_S$ values. Using a similar procedure as described earlier for the $\gamma$-Fe$_2$O$_3$ NPs, $K_{sh}$ and $d$ were determined from the fitting of the overall data using Eq. (4). The variations predicted by Eq. (3) and Eq. (4) are shown in Fig. 5 to emphasize the fact that Eq. (3) is also inadequate to explain the $K_{eff}$ vs. $D$ variation in very small antiferromagnetic NiO NPs. Instead the CSSL model provides a satisfactory description of the observed size dependence of $K_{eff}$ in NiO NPs similar to results for the $\gamma$-Fe$_2$O$_3$ NPs shown in Fig. 4.

Next, the validity of the CSSL model is tested for the ferromagnetic Ni NPs using the data available in the literature [11, 14, 28-30]. Again exactly the same guidelines and procedures regarding the IPI were employed in selecting the data as described earlier in the case of NPs of NiO and $\gamma$-Fe$_2$O$_3$. The plot of the $K_{eff}$ vs. $1/D$ data for Ni NPs is shown in Fig. 6 for an adequate number of the particle sizes. Once again we observe that Eq. (3) is unable to fit the data of the smaller Ni NPs, whereas the CSSL model satisfactorily explains the size dependence of $K_{eff}$ for all considered sizes of Ni NPs. This is due to the fact that compared to the $K_b$ and $K_S$ values, the shell anisotropy $K_{sh}$ term dominates the net $K_{eff}$ for smaller NPs as discussed in Ref. [10].

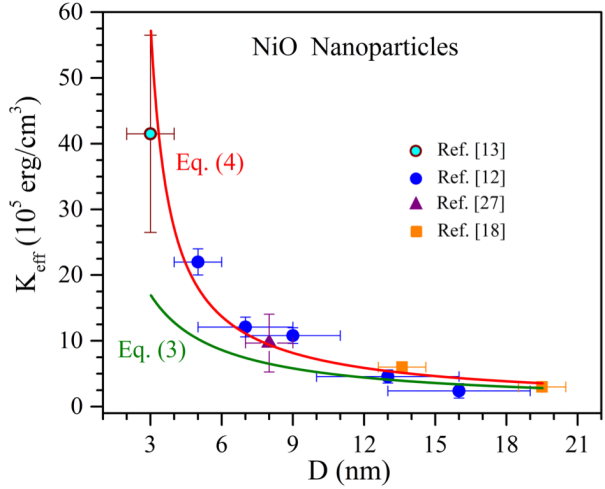

Figure 5: Variation of $K_{eff}$ with particle size $D$ for NiO NPs. Different symbols represent data reported from different groups as cited in the inset. The green line is the best fit corresponding to Eq. (3) while red line represents the best fit corresponding to Eq. (4) with the following magnitudes of the fitting parameters: $K_b = 3.0 \times 10^4$ erg/cm$^3$, $K_S = 0.067$ erg/cm$^2$, $K_{sh} = 3.9 \times 10^5$ erg/cm$^3$, $d = 0.85$ nm.

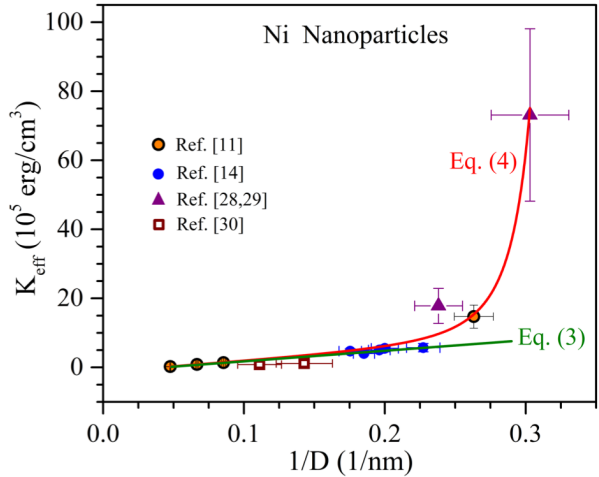

Figure 6: Variation of $K_{eff}$ with $1/D$ for Ni NPs. Different symbols represent data reported from different groups as cited in the inset. The green line is the best fit corresponding to Eq. (3) while red line represents the best fit corresponding to Eq. (4) with the following magnitudes of the fitting parameters: $|K_b| = 1.04 \times 10^5$ erg/cm$^3$, $K_S = 0.05$ erg/cm$^2$, $K_{sh} = 11556$ erg/cm$^3$, $d = 1.45$ nm.

## VI. CONCLUSIONS

In this paper, it has been shown that the size-dependence of $K_{eff}$ in NPs of ferrimagnetic γ-Fe$_2$O$_3$, ferromagnetic Ni and antiferromagnetic NiO is adequately described by the CSSL model proposed here which is described by Eq. (4). This model is an extension of the often used model represented by Eq. (3) by including an additional term due to the spins in a shell of thickness $d$. As discussed in Ref. 10, the contribution from this additional term becomes important only for the smaller particles typically below about 5 nm. The validity of this CSSL model shown here for three different magnetic systems mentioned above suggests that this model should also be applicable for NPs of other magnetic systems in which data of $K_{eff}$ vs $D$ becomes available over a large enough size range without the interference of the interparticle interactions.

ACKNOWLEDGEMENTS

SS acknowledges the support of the Jefimenko Fellowship and Robert T. Bruhn Research Award at West Virginia University. KLP thanks the U.S. National Science Foundation (Grant # DGE-1144676) for financial support of this research. The use of the West Virginia University Shared Research Facilities is also acknowledged.

REFERENCES

[1] Q. A. Pankhurst, N.T.K. Thanh, S. K. Jones, and J. Dobson, "Progress in applications of magnetic nanoparticles in biomedicine", J. Phys. D: Appl. Phys. 42, no. 22 (2009): 224001.
[2] S. P. Gubin, ed. "Magnetic nanoparticles", John Wiley & Sons, 2009.
[3] D. Fiorani, ed. "Surface Effects in Magnetic Nanoparticles", Springer 2005.
[4] N.T. K. Thanh, ed. "Magnetic Nanoparticles: From Fabrication to Clinical Applications", CRC Press, Bocce Raton, FL, 2012.
[5] S. D. Bader; "Colloquium: Opportunities in nanomagnetism", Rev. Mod. Phys. 3, 78(1) 2006.
[6] J. L. Tholence, "On the frequency dependence of the transition temperature in spin glasses", Solid State Commun. 35, 2 (1980): 113-117.
[7] J. L. Dormann, L. Bessais, and D. Fiorani, "A dynamic study of small interacting particles: superparamagnetic model and spin-glass laws", J. Phys. C: Solid State Phys.21, no. 10 (1988): 2015.
[8] M. S. Seehra, and K. L. Pisane, "Relationship between blocking temperature and strength of interparticle interaction in magnetic nanoparticle systems", J. Phys. Chem. Solids 93, 79 (2016).
[9] Kelly Pisane, "Effects of Size and Size Distribution on the Magnetic Properties of Maghemite Nanoparticles and Iron-Platinum Core-Shell Nanoparticles", PhD diss., West Virginia University, 2015.
[10] K. L. Pisane, S. Singh, and M. S. Seehra, "Unusual enhancement of effective magnetic anisotropy with decreasing particle size in maghemite nanoparticles", Appl. Phys. Lett. 110 (22), 222409 (2017).
[11] V. Singh, M. S. Seehra, and J. Bonevich, "ac susceptibility studies of magnetic relaxation in nanoparticles of Ni dispersed in silica", J. Appl. Phys. 105, no. 7 (2009): 07B518.
[12] H. Shim, P. Dutta, M. S. Seehra, and J. Bonevich, "Size dependence of the blocking temperatures and electron magnetic resonance spectra in NiO nanoparticles." Solid State Commun. 145, no. 4 (2008): 192-196.
[13] E. Winkler, R. D. Zysler, M. Vasquez Mansilla, and D. Fiorani, "Surface anisotropy effects in NiO nanoparticles", Phys. Rev. B 72, 132409 (2005).
[14] S. H. Masunaga, R. F. Jardim, P. F. P. Fichtner, and J. Rivas, "Role of dipolar interactions in a system of Ni nanoparticles studied by magnetic susceptibility measurements", Phys. Rev. B 80, 184428 (2009).
[15] G. F. Goya, F. C. Fonseca, R. F. Jardim, R. Muccillo, N. L. V. Carreno, E. Longo, and E. R. Leite, "Magnetic dynamics of single-domain Ni nanoparticles." J. Appl. Phys. 93, no. 10 (2003): 6531-6533.
[16] F. Bødker, S. Mørup, and S. Linderoth, "Surface effects in metallic iron nanoparticles", Phys. Rev. Lett, 72, 282 (1994).
[17] R. Yanes, O. Chubykalo-Fesenko, H. Kachkachi, D. A. Garanin, R. Evans, and R. W. Chantrell, "Effective anisotropies and energy barriers of magnetic nanoparticles with Néel surface anisotropy", Phys. Rev. B 76, 064416 (2007).

[18] M. P. Proenca, C. T. Sousa, A. M. Pereira, P. B. Tavares, J. Ventura, M. Vazquez, and J. P. Araujo, "Size and surface effects on the magnetic properties of NiO nanoparticles", Phys. Chem. Chem. Phys. 13(20):9561-7 (2011).

[19] H. Kachkachi, M. Nogues, E. Tronc and D.A. Garanin, "Finite-size versus surface effects in nanoparticles", J. Magn, Magn. Mater. 221, 158 (2000).

[20] K. L. Krycka, R. A. Booth, C. R. Hogg, Y. Ijiri, J. A. Borchers, W. C. Chen, S. M. Watson, M. Laver, T. R. Gentile, L. R. Dedon, S. Harris, J. J. Rhyne, and S. A. Majetich, "Core-shell magnetic morphology of structurally uniform magnetite nanoparticles", Phys. Rev. Lett. 104, 20 (2010): 207203.

[21] K. L. Krycka, J. A. Borchers, R. A. Booth, Y. Ijiri, K. Hasz, J. J. Rhyne, and S. A. Majetich, "Origin of surface canting within $Fe_3O_4$ nanoparticles", Phys. Rev. Lett. 113, 14 (2014): 147203.

[22] K. Nadeem, H. Krenn, T. Traussnig, R. Wurschum, D.V. Szabo and I. Letofsky-Papst, "Effect of dipolar and exchange interactions on magnetic blocking of maghemite nanoparticles", J. Magn. Magn. Mater. 323, 1998 (2011).

[23] D. Fiorani, A.M. Testa, F. Lucari, F. D'Orazio, and H. Romero, "Magnetic properties of maghemite nanoparticle systems: surface anisotropy and interparticle interaction effects", Physica B, 320, 122 (2002).

[24] P. Demchenko, N. Nedelko, N. Mitina, S. Lewińska, P. Dłużewski, J.M. Greneche, S. Ubizskii, S. Navrotskyi, A. Zaichenko and A. Ślawska-Waniewska, "Collective magnetic behavior of biocompatible systems of maghemite particles coated with functional polymer shells", J. Magn. Magn Mater. 379, 28 (2015).

[25] F. Gazeau, J. C. Bacri, F. Gendron, R. Perzynski, Yu. L. Raikher, V. I. Stepanov and E. Dubois, "Magnetic resonance of ferrite nanoparticles: evidence of surface effects", J. Magn. Magn. Mater. 186, 175 (1998).

[26] S.S. Laha, R.J Tackett and G. Lawes, "Interactions in $\gamma$-$Fe_2O_3$ and $Fe_3O_4$ nanoparticle systems", Physica B 448, 69 (2014).

[27] F. Bødker, M. F. Hansen, C. B. Koch, and S. Mørup, "Particle interaction effects in antiferromagnetic NiO nanoparticles", J. Magn, Magn. Mater., 221(1), pp.32-36 (2000).

[28] G. F. Goya, F. C. Fonseca, R. F. Jardim, R. Muccillo, N. L. V. Carreno, E. Longo, and E. R. Leite, "Magnetic dynamics of single-domain Ni nanoparticles", J. Appl. Phys. 93, 10 (2003): 6531-6533.

[29] F. C. Fonseca, G. F. Goya, R. F. Jardim, R. Muccillo, N. L. V. Carreño, E. Longo, and E. R. Leite, "Superparamagnetism and magnetic properties of Ni nanoparticles embedded in $SiO_2$", Phys. Rev. B 66, 104406 (2002).

[30] F. C. Fonseca, R. D. Jardim, M. T. Escote, P. S. Gouveia, E. R. Leite, and E. Longo, "Superparamagnetic Ni: $SiO_2$–C nanocomposites films synthesized by a polymeric precursor method", Journal of Nanoparticle Research; Feb 1;13(2):703-10 (2011).